\documentclass[12pt]{article}

\usepackage{latexsym,amsfonts,bm,epsfig,amsmath,amsthm,thmtools,amssymb}
\usepackage{mathtools}

\usepackage{float} 
\usepackage{booktabs} 
\usepackage{graphicx} 
\usepackage[margin=1cm]{caption} 
\usepackage{subfigure}
\usepackage{authblk}

\usepackage[comma]{natbib}
\usepackage{url}
\usepackage{xcolor}
\newcommand{\cut}{\mathrm{cut}}

\begin{document}


\title{Bayesian inference for misspecified generative models}

\author{David J. Nott,$^1$ Christopher Drovandi,$^{2,3}$ and David T. Frazier$^4$
\affil{$^1$Department of Statistics and Data Science, National University of Singapore, Singapore; email: standj@nus.edu.sg}
\affil{$^2$School of Mathematical Sciences, Queensland University of Technology,
 Australia}
\affil{$^3$Centre for Data Science, Queensland University of Technology,
	Australia}
\affil{$^4$Department of Econometrics and Business Statistics, Monash University, Australia}}

\maketitle

\begin{abstract}
Bayesian inference is a powerful tool for combining information in 
complex settings, 
a task of increasing importance in modern applications.  However, Bayesian
inference with a flawed model can produce unreliable conclusions.  
This review discusses approaches to performing Bayesian inference 
when the model is misspecified, where by misspecified we mean that
the analyst is unwilling to act as if the model is correct.  
Much has been written about this topic, and in most cases we do not believe that a conventional
Bayesian analysis is meaningful when there is serious model 
misspecification.   Nevertheless, in some cases it is possible to use a well-specified model to give meaning to a Bayesian analysis of a misspecified model and we will focus on such cases.   Three main classes of methods 
are discussed - restricted likelihood methods, which use a model based on a non-sufficient summary of the original data;  modular inference
methods which use a model constructed from coupled submodels and
some of the submodels are correctly specified;  and the use of a reference
model to construct a
projected posterior or predictive 
distribution for a simplified model
considered to be useful for prediction or interpretation.  
 
\end{abstract}

\section{Introduction}

The advantages of Bayesian inference are well known for problems that
involve combining information.  Expressing uncertainty probabilistically 
and building hierarchies to ``borrow strength" across
related problems is an attractive strategy in many complex statistical
applications.  However, a conventional 
Bayesian analysis may lose its meaning when the model is misspecified;  see 
Walker (2013) and the accompanying discussion for a broad overview
of the issues.

In this review, we discuss how
a meaningful Bayesian analysis can be given
for a misspecified model in some cases.   First we must clarify
what we mean by misspecification.
It is convenient to call a statistical model correctly
specified if we are happy to act as if it is correct.  
\cite{kass+w96} explained a similar point of view using the 
analogy of a clock that keeps good though not perfect time;  it is convenient to say that the clock is correct if any inaccuracy 
does not matter for the things we use it for.  The
qualification ``for what we use it for" is often implicit.  
In a similar spirit, \citet[Chapter 8]{ohagan+f04} talk about a
Bayesian model specification being ``good enough" after exploring
sensitivity of Bayesian inferences using the tools of Bayesian model
criticism.  Similar sentiments have been expressed by many other
authors.  Throughout the rest of this review, a correctly specified model
means one that is correct in the above sense - it need not perfectly
represent the information available.  
In contrast, a misspecified model is one for which
we are not happy to act as if it is correct, and in such cases  
we would not be satisfied with a conventional Bayesian analysis using 
the postulated model.  

Throughout this review we consider generative probabilistic models
of the type used in most Bayesian analyses.  This restriction
makes the scope of our review manageable, but means some
interesting topics are left out.  In particular, we do not say much about
Bayesian methods involving Gibbs posteriors (e.g. \citealp{zhang06,zhang06b,jiang+t08}), 
PAC-Bayes methods \citep{shaw-taylor+w97,catoni07,alquier21},
coherent Bayesian
updating via loss functions \citep{bissiri2016general,jewson+r22}, 
or Bayes linear 
methods \citep{goldstein+w07}, among other worthy ideas related 
to our subject.  We also do not directly discuss the
important topic of Bayesian
model criticism.  If a model is
misspecified, it is natural to use diagnostic methods to understand
and improve the model until we judge it to be adequate for some chosen task.  
Useful overviews of Bayesian model criticism 
are given by \cite{gelman+ms96}, 
\cite{bayarri+b00} and \cite{evans15}.  
The existing work
deals not only with checking the specification of the likelihood, but also with
detecting conflicting information in different parts of the model 
\citep{presanis+osd13} such as between likelihood and prior \citep{evans+m06}.
Some of the modified Bayesian analyses we discuss here can 
play a role in Bayesian model criticism.  Here we assume 
that a decision has been made to work with a 
misspecified model for reasons of practicality.  
For example, a more realistic model might require information that is difficult or expensive to obtain or might be too complex to interpret easily.  

Can a modified 
Bayesian analysis of a misspecified model be meaningful?  The answer to this question might be no in many cases, 
but sometimes more can be said.  We focus on cases where 
there is a well-specified model related to
the misspecified one, which
can be used to give meaning to a modified Bayesian 
analysis for the misspecified model.  
We focus on three topics - restricted likelihood methods, 
which reduce the full data to a summary statistic and model the summary;  
Bayesian modular inference methods which use a model
built from coupled submodels in which some of the submodels
are correctly specified;  and the use of a correctly specified reference
model for constructing projected
posterior or predictive 
distributions for a simplified model which may be more useful
for some purposes.  The idea of using a well-specified model to 
give meaning to a Bayesian analysis of a misspecified model
is related to the ``${\mathcal M}$-completed" view of Bayesian model comparison
\citep[Section 6.1]{bernardo+s09}.  However, we will mostly not discuss
model comparison, except in relation to projection methods in Section 4.  
  

The rest of the paper is structured as follows.  Section 2 describes
the restricted likelihood approach to dealing with misspecification, the computational and statistical challenges 
of the approach, and connections with likelihood-free inference methods.  
Section 3 then considers modular Bayesian inference, with a particular emphasis
on so-called cutting feedback methods.  Section 4 discusses using a 
correctly specified reference model to derive a projected 
posterior or predictive distribution for a simplified model.  Section 5 discusses future challenges in the areas we have
considered.  

\section{Restricted likelihood methods}

The first topic of our review 
is the restricted likelihood approach to robust
Bayesian inference.  The recent
paper by \cite{Lewis2021} and its discussion gives a comprehensive
summary of the method and related literature.  
The main idea is to reduce the available data to an 
insufficient summary statistic which discards  
information that cannot be matched
under the assumed model.  If the 
summary statistic is carefully chosen, a misspecified model for the full
data can induce a model which is an acceptable approximation
to the data generating process for the summary statistic.  
Connecting with the theme of this review, the
summary statistic model can be regarded as correct, and this can
be used to give a meaningful 
Bayesian analysis related to the original misspecified
model for the full data.   
Challenges of the approach are the choice of summary statistics, the
differing meaning of the parameter in the full and summary statistic
models when there is misspecification, 
and the difficulty of computation when the summary statistic
likelihood is not computationally tractable.  Restricted likelihood is related to
other approaches to robust Bayesian inference which involve discarding
information, such as data coarsening \citep{miller+d19}, power
likelihood (e.g. \citealp{grunwald12,grunwald+v17}), minimally
informative likelihood \citep{yuan+c99} and 
Bayesian data selection \citep{weinstein+m23}, although these connections
will not be discussed further here.

We discuss robust Bayesian estimation of regression models, similar
to \cite{Lewis2021}. \cite{Lewis2021} focus specifically on linear regression, and in the discussion of their paper \citet{Drovandi2021} explore likelihood-free inference methods as an alternative approach to computation. These methods are applicable whenever data can be simulated from the model, even if likelihood computation is infeasible. \citet{Drovandi2021} found that the use of likelihood-free methods produced similar results to those of \citet{Lewis2021}, albeit with higher computational costs. However, unlike the approach of \citet{Lewis2021}, likelihood-free methods are not restricted to linear models and have the potential to generalize to more complex regression problems.

As a simple motivating example beyond the linear regression
setting, we consider 
a generalised linear model (GLM) for the damaged carrots dataset from \citet{Phelps1982}, where the response variable is the number of carrots showing insect damage in a soil experiment.  The experiment considered eight dose levels of insecticide and three blocks.  Let $y_i$ be the number of affected carrots out of $n_i$ carrots for the $i$th observation.  It is assumed that $y_i \sim \mbox{Binomial}(n_i, p_i)$ where
\begin{align*}
	\mbox{logit}(p_i) &= \beta_0 + \beta_1 \cdot \log(\mbox{dose}_i) + \beta_2 \cdot \mbox{block2}_i + \beta_3 \cdot \mbox{block3}_i,
\end{align*}
where block2$_i$ and block3$_i$ are binary variables indicating 
whether the $i$th observation
is from block2 or block3, respectively.  The intercept $\beta_0$ 
is the mean response for an observation in the first block with
zero log dose.   The analysis in \citet{Cantoni2004} indicates that the 14th observation is an extreme outlier with respect to the assumed model, and thus robust techniques are of potential interest for this dataset.  Later we will implement
likelihood-free inference approaches to restricted likelihood for this 
example.
 
\subsection{The restricted likelihood approach}

Suppose that $\theta$ is a parameter in a parametric model for
data $y$.  The density of $y$ given $\theta$ is denoted $f(y\mid\theta)$, 
and the density of a data summary $S(y)$ will be denoted
$f(S(y)\mid\theta)$.  We consider Bayesian inference with prior
density $\pi(\theta)$.  
Bayesian restricted likelihood (BRL) proceeds by targeting the posterior conditioned on $S(y)$, as opposed to the full dataset $y$:
\begin{align}
	\pi(\theta\mid S(y)) \propto f(S(y)\mid\theta)\pi(\theta). \label{eq:partial_posterior}
\end{align}
The motivation for BRL is to discard features of 
$y$ that cannot be matched under the assumed model and are not relevant
to the analysis.  The full likelihood can be decomposed as
\begin{align*}
	f(y\mid\theta) &= f(S(y)\mid\theta)f(y\mid\theta,S(y)),
\end{align*}  
where $f(y\mid\theta,S(y))$ is the conditional density of $y$ given $\theta,S(y)$.  
There is a loss of information in replacing $f(y\mid\theta)$ with $f(S(y)\mid\theta)$ for Bayesian inference 
when $S(y)$ is not sufficient.  However, in the setting of model misspecification, it may be desirable to discard information 
in the data that we know the model cannot replicate.  One example given in \citet{Lewis2021} is that of deliberate censoring to limit the influence of small or large observations on an analysis.  Considering a summary statistic $S(y)$ of the same dimension as $y$ and
writing $S(y_i)$ for the $i$th element of $S(y)$, the censoring approach sets $S(y_i) = t_1$ if $y_i < t_1$, $S(y_i) = t_2$ if $y_i > t_2$ and $S(y_i) = y_i$ otherwise.  \citet{Lewis2021} provide a literature review on different types of conditioning statistics considered in various scenarios, such as the use of rank and robust statistics.  The authors also state some theoretical results suggesting that the posterior conditioned on $S(y)$ can resemble the asymptotic distribution of the conditioning statistic, under certain conditions.

\citet{Lewis2021} focus on robust Bayesian estimation of linear models:
\begin{align*}
	y_i = z_i^\top \beta + \epsilon_i, \quad \mbox{for } i = 1,\ldots,n,
\end{align*}
where $z_i \in \mathbb{R}^p$ is the vector of covariates for the $i$th observation, $\beta \in \mathbb{R}^p$ is the vector of regression coefficients and $\epsilon_i$ is the $i$th residual, assumed to have mean 0 and scale $\sigma$.  The model parameter $\theta = (\beta^\top, \sigma^2)$ is allocated a prior distribution $\pi(\theta)$.
\citet{Lewis2021} consider summary statistics that are estimates of the model parameters obtained when fitting the regression model using a robust frequentist estimation technique (e.g.\ \citealp{Huber2009}), such as least median squares, least trimmed squares and M-estimators based on, for example, Huber's \citep{Huber1964} or Tukey's \citep{Beaton1974} loss.  

Although BRL addresses the robustness issue, it introduces a computational challenge.  Even when the likelihood function $f(y\mid\theta)$ is feasible to evaluate, the summary statistic likelihood $f(S(y)\mid\theta)$ can be intractable.  
The approach of \citet{Lewis2021} considers an augmented MCMC algorithm in which a data replicate denoted $x$ is considered.  We do not give an introduction to MCMC methods here, but if a Gibbs sampling algorithm were
feasible one could iteratively sample from 
the full conditionals $\pi(\theta\mid x, S(x) = S(y))=\pi(\theta|x)$ and $f(x\mid\theta, S(x) = S(y))$.  More generally, the Gibbs steps can be replaced by Metropolis-Hastings updates.  
For $\pi(\theta|x)$, the posterior if data $x$ was observed, standard approaches can be used to 
design a valid Metropolis-Hastings update.  Designing
the update for  $f(x\mid\theta,S(x)=S(y))$ requires
more ingenuity.  First, define the space of potential datasets that have a summary statistic matching the observed one, $\mathcal{A} = \{x \in \mathbb{R}^n | S(x) = S(y)\}$.  Define the proposal density for $x \in \mathcal{A}$ as $q(x\mid\theta)$. The Metropolis-Hastings ratio for the proposal, $x^* \sim q(\cdot\mid\theta)$, is given by
\begin{align*}
	R &= \frac{f(x^*\mid\theta) q(x\mid\theta)}{f(x\mid\theta) q(x^*\mid\theta)}.
\end{align*}
Theorem 3.1 of \citet{Lewis2021} considers a particular form of the summary statistic and shows that any dataset $z \in \mathbb{R}^n$ with statistic $S(z)$ can be transformed into a dataset $x$ with statistic $S(y)$ by
\begin{align*}
	x &= \frac{s(X,y)}{s(X,z)}z + X\left(  b(X,y) - b\left(  X ,  \frac{s(X,y)}{s(X,z)}z\right)  \right),
\end{align*}
where $X$ is the design matrix, $b(X,\cdot)$ is a vector of summary statistics related to the regression coefficients and $s(X,\cdot)$ is the summary statistic related to the scale parameter.  The initial dataset $z$ is drawn from a known distribution, so that the proposal density for $x$ can be obtained by standard application of the result for transformation of random variables.

\subsection{Connections with likelihood-free methods}

Although \citet{Lewis2021} provide an elegant solution for avoiding evaluation of the summary statistic likelihood in the linear model setting, such an approach may not be feasible in more complicated settings, such as in generalised linear models.  In the rejoinder of their discussion paper, \citet{Lewis2021} note that ``...\ straightforward use of our techniques will
break down for models with enough complexity."

Fortunately, there are several so-called likelihood-free, or simulation-based, Bayesian inference methods (see \citet{Sisson2018} and \citet{cranmer+bl20} for reviews) that have been developed in the literature for estimating likelihoods of summary statistics based on model simulations.   Such methods are useful when estimating the parameters of complex stochastic processes, which are feasible to simulate but for which the corresponding likelihood function is too computationally expensive to evaluate.  Therefore, there is potential to harness likelihood-free methods for BRL beyond the linear model setting.  

An earlier BRL approach in \citet[chap.\ 4]{Lewis2012} is reminiscent of approximate Bayesian computation (ABC), a popular likelihood-free method.   
For a given value of $\theta$, simulate $m$ independent datasets, each of size $n$.  Write $x_i$ for the $i$th simulated dataset, and $S(x_i)$ for the corresponding summary statistic.
ABC estimates the summary statistic likelihood by
\begin{align}
f_\epsilon({S(y)\mid\theta }) = \frac{1}{m} \sum_{i=1}^m K_\epsilon(\rho\{S(y), S(x_i)\}), \label{eq:ABC_likelihood}
\end{align}
where $\rho\{S(y), S(x)\}$ measures the discrepancy between observed and simulated summaries and $K_\epsilon(\cdot)$, $\epsilon>0$, is a kernel that allocates higher weight to smaller $\rho$.  The kernel bandwidth $\epsilon$ is often referred to as the tolerance in the ABC literature.  $f_\epsilon(S(y)\mid\theta)$ is effectively a kernel density estimate of the summary statistic likelihood.  \citet[chap.\ 4]{Lewis2012} consider a more general kernel density estimate with a matrix bandwidth, however this is not often considered in ABC.

Given the likelihood estimator in \eqref{eq:ABC_likelihood}, approximate simulation from the target in \eqref{eq:partial_posterior}, often referred to as the partial posterior in the likelihood-free context, can proceed by using an importance sampling or Markov chain Monte Carlo based algorithm (see \citet{Sisson2018a} for a review).
The non-parametric ABC likelihood estimator in \eqref{eq:ABC_likelihood}
can be computationally inefficient, particularly when the summary statistic is high-dimensional.  An alternative approach is Bayesian synthetic likelihood (BSL, \citet{wood10,Price2018,Frazier2022}) which employs a parametric Gaussian approximation instead,
$$
f_A({S(y)\mid\theta }) = \mathcal{N}\left(S(y);\mu_m(\theta),\Sigma_m(\theta)\right),
$$
where $\mu_m$ and $\Sigma_m$ are calculated from the $m$ model simulations using sample moments
\begin{flalign*}
	\mu_m(\theta)&=\frac{1}{m}\sum_{i=1}^{m}S(x_i), \\
	\Sigma_m(\theta)&=\frac{1}{m}\sum_{i=1}^{m}\left(S(x_i)-\mu_m(\theta)\right)\left(S(x_i)-\mu_m(\theta)\right)^{\top}.
\end{flalign*}
The parametric restriction leads to computations scaling better with summary statistic dimension \citep{Frazier2022}, at the expense of the strong Gaussian assumption.  BSL may be an attractive option for general BRL problems, since popular robust estimators such as M-estimators are asymptotically normal under mild conditions \citep{van2000asymptotic}.  If there is concern about assuming normality with small samples, then the semi-parametric extension of \citet{An2020} could be employed.

ABC and BSL have been studied recently under model misspecification.  Most work considers a strong type of misspecification called incompatibility \citep{Marin2014}, where the expected value of the summary statistic under the true data generating process and of the assumed model do not coincide for any value of $\theta$.  Loosely, this means that we cannot obtain close matches between $S(x)$ and $S(y)$ for any value of $\theta$.  
However, as we discuss next, even if the summary statistic model is not
correctly specified, it may be possible to consider a model expansion
that can result in an adequate specification, and which can be insightful
about which of the summary statistics can and can't be matched under
the assumed model.  

\citet{Frazier2021a} develop a solution to the incompatibility problem for BSL, which is further analysed theoretically in \citet{Frazier2021}.  The approach involves incorporating an auxiliary parameter, $\Gamma \in \mathbb{R}^d$ such that $\Gamma = (\gamma_1,\ldots,\gamma_d)^\top$, where $\gamma_i$ corresponds to the $i$th summary statistic.   The approach of \citet{Frazier2021a} adjusts the mean or inflates the variance of the synthetic likelihood so that the observed summary has reasonable support in the expanded model in the region of the pseudo-true parameter value.  The expanded model has a parameter with dimension greater than that of the summary statistic since $\dim((\theta,\Gamma)^\top) = d + d_\theta$.    \citet{Frazier2021a} impose a prior distribution on $\Gamma$ that favours compatibility to regularise the model.  Each component of the prior for $\Gamma$ has a heavy tail so that it can ``soak up" the misspecification for the incompatible summary statistics.  Thus, the method is able to identify the incompatible summaries, and at the same time, mitigate the influence of these statistics on the inference.  The posterior for $\Gamma$ is the same as its prior under compatibility, so that incompatibility can be detected by departures from the prior.  For the variance inflation approach, the regularised BSL covariance is given by
\begin{align*}
\Sigma_m(\theta, \Gamma) = \Sigma_m(\theta)+\Sigma_m^{1 / 2}(\theta) \operatorname{diag}\left\{\gamma_1, \ldots, \gamma_d\right\} \Sigma_m^{1 / 2}(\theta),
\end{align*}
while for mean adjustment the BSL mean component is given by
\begin{align*}
	\mu_m(\theta,\Gamma) &=	\mu_m(\theta)+ \sigma_m(\theta) \circ \Gamma,
\end{align*}
where $\circ$ denotes element-wise multiplication and $\sigma_m(\theta)$ is the vector of standard deviations of the summary statistics.



Another highly relevant approach for conducting posterior inference based on misspecified generative models is the Q-posterior developed in \citet{Frazier2023}.  Here, a general methodology is proposed to obtain posteriors that provide (asymptotically) accurate frequentist coverage of the pseudo-true parameter value in misspecified models.  Like \citet{Lewis2021}, the method is applicable in cases where the likelihood is tractable, and it is also based on conditioning on a particular choice of the summary statistic.

Define $m(\theta) = -\nabla_\theta \log p(y\mid\theta)$ as the gradient of the negative log-likelihood.  \citet{Frazier2023} consider performing inferences with $n^{-1} m(\theta)$ as the summary statistic.  Defining $W(\theta)$ as a consistent estimator of $\mbox{Cov}(m(\theta)/\sqrt{n})$ and noting that the expected value of $n^{-1} m(\theta)$ under the assumed model is a vector of zeros, \citet{Frazier2023} propose the following approximate posterior that uses a Gaussian model for $m(\theta)$ in a similar spirit to synthetic likelihood
\begin{align*}
	\pi_Q(\theta\mid y) & \propto |W_n(\theta)|^{-1/2} \exp\left\{- \frac{1}{2}\frac{m(\theta)^\top}{\sqrt{n}}W(\theta)^{-1}\frac{m(\theta)}{\sqrt{n}}  \right\}  \pi(\theta),
\end{align*}  
which is referred to as the Q-posterior in \citet{Frazier2023}.  The authors show that credible regions produced by the Q-posterior are accurate in the sense they have the correct frequentist coverage.  This result can still hold even when the likelihood involves an intractable integral, for example in mixed effects models.  
The Q-posterior falls within the framework of generalised posteriors in the spirit of \cite{bissiri2016general}, but conveniently does not require calibration of
any scaling parameters as is typically the case in generalised Bayesian analyses.  



\subsection{Generalised Linear Model Example}

We return to the logistic regression example for the damaged carrots dataset.  Three Bayesian inference approaches are compared.  Firstly, we consider the true posterior based on the actual likelihood, which may not be robust to the outlier.  We also consider BRL, with the summaries set as the robust estimators implemented in the \texttt{glmrob} function within the \texttt{robustbase} R package \citep{Maechler2022}.  We use the Huber quasi-likelihood estimator of \citet{Cantoni2001} with the tuning constant $c$ set in Huber's $\psi$ function to be $c=1.2$.   The conditions required for Theorem 3.1 in \citet{Lewis2021} for exact conditioning are developed for the linear model, and the extension of BRL to nonlinear models such as GLMs does not seem feasible due to the computational challenges associated with producing exact draws from the corresponding posterior.  Thus, we adopt a simulation-based approach, namely BSL with $m=20$, to approximate the restricted or partial posterior.  We refer to this as Bayesian restricted synthetic likelihood (BRSL).  We also implement the Q-posterior approach of \citet{Frazier2023}.  

In all cases, we use 50,000 iterations of MCMC with a multivariate normal random walk proposal.  For BRSL we use the estimate from \texttt{glmrob} as the starting value for the chain and set the random walk covariance as the covariance of the parameter estimates from  \texttt{glmrob}.  For the true posterior and Q-posterior we use the starting value from the classical \texttt{glm} fit to the data and set the random walk covariance as the covariance of the parameter estimates from \texttt{glm}.  Since we use good starting values for the MCMC chains we do not use any burn-in.   The BRSL approach is more computationally intensive as it involves model simulation.  To implement MCMC for BRSL we use the \texttt{BSL} R package \citep{An2022}, and for MCMC sampling of the true and Q posteriors we use the \texttt{mcmc} R package \citep{Geyer2020}.

\begin{figure}[h]
	\centering{\includegraphics[scale=0.7]{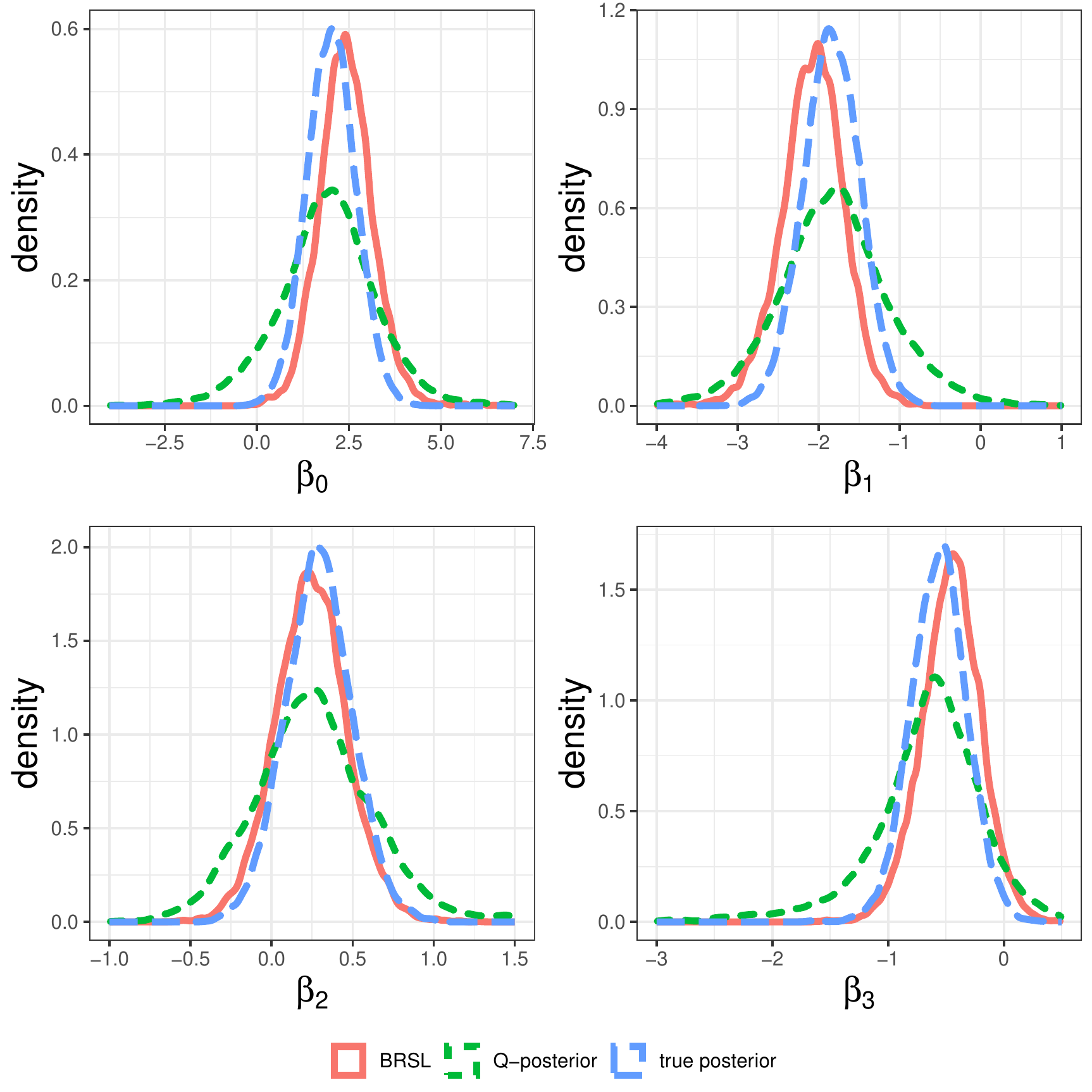}}
	\caption{Posterior approximations for the damaged carrots dataset.}\label{fig:results_glm}
\end{figure}

The estimated univariate posteriors are shown in Figure \ref{fig:results_glm} (based on thinning each chain by a factor of 10).  Here it appears that the Q-posterior and BRSL approximations accommodate the misspecification in different ways.  The Q-posterior is centred closely on the same location as the true posterior, but with thicker tails in order to more accurately quantify the uncertainty of the pseudo-true parameter value, which is very close to the MLE.  In contrast, the BRSL posterior mode is shifted slightly by reducing the influence of the outlying observation, and has a posterior variance similar to the true posterior.

\section{Bayesian modular inference}

The second topic of our review is Bayesian modular inference, with
a focus on ``cutting feedback" methods
\citep{liu+bb09,jacob+mhr17}.  Bayesian modular inference is 
concerned with
multi-modular models, which are constructed by coupling simple submodels
together.  Usually the different submodels (called modules)
have their own sources of data, and there may be both 
shared and module specific parameters.  
Sometimes there 
is more confidence in the specification of some modules than others.  
In these cases, it can be desirable to conduct a modified version of Bayesian
inference called ``cutting feedback" \citep{lunn+bsgn09,plummer15} which cuts
the feedback from suspect modules to
achieve insensitivity to misspecification.  We can use Bayesian inference
restricted to some of the correctly specified modules to achieve more 
meaningful inference in the larger misspecified model.
The challenges of applying the method include giving general definitions
of modules and cut posterior distributions, deciding whether a full or cut 
posterior distribution is preferable, and cut model posterior computation, 
which is more difficult than conventional Bayesian posterior computation.  

While Bayesian modular inference is a broad term, we discuss
only cutting feedback methods in the rest of this section.  
To explain the main idea, it is helpful to consider a simple two module
system originally discussed in \cite{plummer15}.  We consider a model where
there are two sources of data, given by vectors  $X$ and $Y$.  The
complete data is denoted $Z=(X^\top,Y^\top)^\top$.  The density of 
$Z$ is $f(Z\mid\theta)$, where $\theta$ is a vector of parameters which can
be partitioned as $\theta=(\zeta^\top,\varphi^\top)^\top$.  
The density of $Z$ decomposes as
$$f(Z\mid\theta)=f_1(X\mid\varphi)f_2(Y\mid\zeta,\varphi),$$
so that $X$ and $Y$ are conditionally independent given
the parameters, with the density of $X$ depending
on $\varphi$, and the density of $Y$ depending on $\zeta$ and $\varphi$.  
We consider Bayesian inference with a prior density 
$\pi(\zeta,\varphi)=\pi(\varphi)\pi(\zeta\mid \varphi)$.

This model has a ``multi-modular" form, with 
the first module consisting of $\pi(\varphi)$ and $f_1(X\mid\varphi)$, 
and the second module consisting of $\pi(\zeta\mid\varphi)$ and 
$f_2(Y\mid\zeta,\varphi)$.  
Bayesian inference could be conducted 
for $\varphi$ based on the first module only;  on the other hand, 
given a value for $\varphi$, conditional inference for $\zeta$ can be
conducted based on the second module only.  
The parameter $\varphi$ is global, being shared between the two
modules, while $\zeta$ is a parameter specific to the second module. 
Figure \ref{two-module} gives a graphical
representation, where the red dashed line is a ``cut" with 
nodes to the left of the 
line comprising module 1, and nodes to the right comprising 
module 2.  The meaning of the ``cut" will be explained later.  In Figure 
\ref{two-module}
it has been assumed that
the prior $\pi(\zeta\mid\varphi)$ does not depend on $\varphi$ for 
simplicity.  

In a conventional Bayesian analysis of this model, there is
``feedback'' between the two modules. This feedback ensures that if either module is misspecified, the
misspecification will contaminate inferences for both components. 
Noting that the conditional posterior of $\zeta$ given $\varphi$ 
depends on the data only through $Y$, the joint posterior density is
\begin{flalign}\label{eq:exactpost}
	\pi(\theta \mid Z_{})&= \pi(\zeta \mid Y_{}, \varphi)\pi(\varphi\mid X_{}, Y_{})=\frac{f_2(Y\mid\zeta,\varphi)\pi(\zeta\mid\varphi)}{P(Y\mid\varphi)}\frac{P(Y\mid\varphi)f_1(X\mid\varphi)\pi(\varphi)}{P(X,Y)},
\end{flalign}where $P(Y\mid\varphi)=\int f_2(Y\mid\zeta,\varphi)\pi(\zeta\mid\varphi) d\zeta$ and $P(X,Y)=\int f_1(X\mid\varphi)f_2(Y\mid\zeta,\varphi)\pi(\zeta\mid\varphi )\pi(\varphi)d\zeta d\varphi$.
If $f_2(Y\mid\zeta,\varphi)$ is misspecified, the resulting posterior for $\varphi$ can be impacted, since the definitions of $P(Y\mid\varphi)$ and $P(X,Y)$ both depend on the specification of $f_2(Y\mid\zeta,\varphi)$.  
The marginal posterior density for $\varphi$ is different to the one
obtained from a Bayesian analysis using 
the first module only due to the presence of the ``feedback" term $P(Y\mid\varphi)$.

\subsubsection*{Motivating Example}
To better understand how feedback in one module may adversely impact inferences for parameters in another module, we consider a misspecified normal-normal random effects model discussed in \cite{liu+bb09}. 
We observe data $Z_{ij}$ comprising observations on  $i=1,\dots,N$ groups, with $j=1,\dots,J$ observations in each group, which we assume are generated from the model $Z_{ij}\mid \beta_i,\varphi_i^2\stackrel{iid}{\sim} N(\beta_i,\varphi_i^2)$, with random effects $\beta_i\mid\psi \stackrel{iid}{\sim} N(0,\psi^2)$ and group variance parameters $\varphi_i^2$, $i=1,\dots, N$. The goal of the analysis is to conduct inference on the standard deviation of the random effects, $\psi$, and the residual standard deviation parameters $\varphi=(\varphi_1,\dots,\varphi_N)^\top$.  Below we write $\beta=(\beta_1,\dots, \beta_N)^\top$, and
$\zeta=(\psi,\beta^\top)^\top$.  

Write $\bar{Z}_i=J^{-1}\sum_{j=1}^{J}Z_{ij}$ and $s_i^2=\sum_{j=1}^J (Z_{ij}-\bar{Z}_i)^2$, $i=1,\dots, N$.  
The likelihood for $\zeta,\varphi$ can be written to depend only on the sufficient statistics $\bar{Z}=(\bar{Z}_1,\dots,\bar{Z}_N)^\top$ and $s^2=(s_1^2,\dots,s_N^2)^\top$, where independently for $i=1,\dots, N$, 
\begin{flalign*}
	& \bar{Z}_i\mid\zeta,\varphi \sim N(\beta_i,\varphi_i^2/J),\\&s^2_i|\varphi\sim \text{Gamma}\left(\frac{J-1}{2},\frac{1}{2}\frac{1}{\varphi_i^2}\right).
\end{flalign*}The random effects model can then be written as a two-module system of the form shown in Figure \ref{two-module}, 
where module one depends on $(s^2,\varphi)$, and module two depends on $(\bar{Z},\zeta,\varphi)$.   In the notation shown in the graph, $X=s^2$ and $Y=\bar{Z}$.  Let $\text{Gamma}(x;A,B)$ denote the value
of the $\text{Gamma}(A,B)$ density evaluated at $x$, and 
$N(x;\mu,\sigma^2)$ denote the value of the $N(\mu,\sigma^2)$ density
evaluated at $x$. The first module has likelihood  $f_1(X\mid \varphi)= \prod_{i=1}^{N}\text{Gamma}\left(s_i^2;\frac{J-1}{2},\frac{1}{2}\frac{1}{\varphi_i^2}\right)$, while the second module has likelihood  $f_2(Y\mid \zeta,\varphi)=\prod_{i=1}^{N}N(\bar{Z}_i;\beta_i,\varphi_i^2/J)$,  

If the prior on the random effects $\beta_i\sim N(0,\psi^2)$ is 
misspecified, in the sense that this conflicts with likelihood information, 
then this will corrupt posterior inferences for both $\varphi$ and $\zeta$. 
For example, suppose that, for some value of $i$, the $\beta_i$ term is much larger in
magnitude than the other components of $\beta$.  $\bar{Z}_i$ 
will be correspondingly large in magnitude compared to the other components
of $\bar{Z}$.  The thin-tailed Gaussian prior for the random effects seems inappropriate here, and as shown by \cite{liu+bb09} it may result in 
overshrinkage in estimating the $\beta_i$, 
leading to poor inference about $\varphi_i$, due to the influence of the likelihood term 
$N(\bar{Z}_i;\beta_i,\varphi_i^2/J)$ for the second module.   
This may be particularly problematic when
$N$ is large and $J$ is small.  We return to this example later.

\subsection{Cutting Feedback}

\begin{figure}[h]
	\centering{\includegraphics[width=60mm, height=60mm]{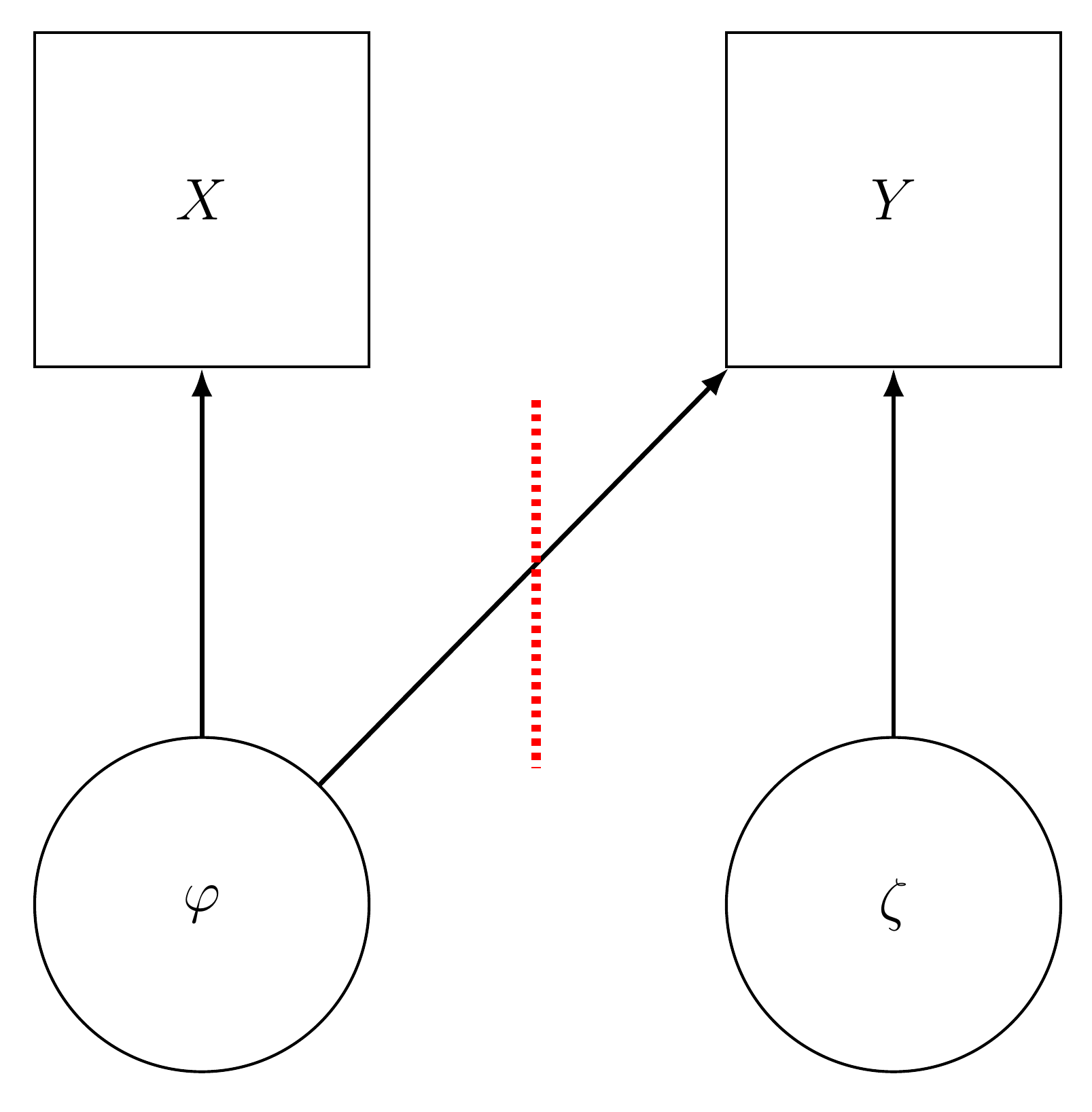}}
	\caption{Graphical structure of the two-module system.  The dashed line indicates the cut.}\label{two-module}
\end{figure}

``Cutting feedback" methods construct a modified 
cut posterior distribution that makes inference in correctly
specified modules insensitive to misspecification in the remaining modules.  
In the context of the two module system, suppose that we are worried about
misspecification of $f_2(Y\mid\zeta,\varphi)$ influencing inferences for $\varphi$.  To make sure this doesn't happen, we can ``cut'' the link between the two modules as shown by the red dashed line in Figure \ref{two-module}.  
This cut involves removing the feedback term $P(Y\mid\varphi)$ from the
numerator on the right-hand side of Equation \eqref{eq:exactpost} and 
renormalizing.  Since this term
represents the influence of the second module on marginal inferences
about $\varphi$, the cut can be interpreted as conducting posterior
inferences for $\varphi$ based on the first module only.  
The $\varphi$ marginal posterior 
density now becomes 
$\pi_{\cut}(\varphi\mid X)=f_1(X\mid\varphi)\pi(\varphi)/\int f_1(X\mid\varphi)\pi(\varphi)d\varphi$, and we can combine this 
with the conditional
posterior density for $\zeta$ given $\varphi$ from
the conventional Bayesian joint posterior to obtain the joint cut posterior density 
\begin{align}
	\pi_{\cut}(\zeta,\varphi\mid X,Y)&=\pi(\zeta \mid Y_{}, \varphi)\pi_{\cut}(\varphi\mid X). \label{cutjoint} \\
	&=\frac{f_2(Y\mid\zeta,\varphi)\pi(\zeta\mid\varphi)}{P(Y\mid\varphi)}\frac{f_1(X\mid\varphi)\pi(\varphi)}{\int f_1(X\mid\varphi)\pi(\varphi)d\varphi}. \nonumber
\end{align}
The cut posterior modifies the conventional posterior density by
replacing $\pi(\varphi\mid X,Y)$ with $\pi(\varphi\mid X)$, and inference about $\varphi$ is
is now unaffected by misspecification in module 2.  
However, uncertainty about $\varphi$ is still propagated 
when making marginal cut posterior inferences about $\zeta$,  due to the conditioning on $\varphi$ in the conditional posterior $\pi(\zeta\mid Y,\varphi)$. 


\subsubsection*{Motivating Example Continued}
Following \cite{liu+bb09}, we can restrict the influence of $\bar{Z}$ on $\varphi$ by using a cut posterior distribution. First consider the 
conventional joint posterior distribution.  We write $\varphi^2$ below
for the square of the vector $\varphi$ elementwise.  Assuming prior independence of components of $\varphi^2$ with $\pi(\varphi_i^2)\propto(\varphi_i^2)^{-1}$, $i=1,\dots, N$, and $\pi(\psi^2\mid\varphi)\propto (\bar\varphi^2/J+\psi^2)^{-1}$, with $\bar\varphi^2=\sum_{i=1}^{N}\varphi_i^2/N$, the marginal posterior distribution for the parameters $\psi^2,\varphi^2$ can be written as (see \citealp{liu+bb09})
\begin{flalign*}
	\pi(\psi^2,\varphi^2\mid \bar{Z},s^2)&=\pi(\psi\mid\bar{Z},\varphi)\pi(\varphi\mid \bar{Z},s^2)\\&\propto\frac{1}{\psi^2+\bar\varphi^2/J}\prod_{i=1}^{N}(\varphi_i^2)^{-\frac{J+1}{2}}\exp\left\{-\frac{J\cdot s_i^2}{2\varphi_i^2}\right\}\times \\
	& \hspace{25mm} \frac{1}{(\psi^2+\varphi_i^2/J)^{1/2}}\exp\left\{-\frac{\bar{Z}_i^2}{2(\psi^2+\varphi_i^2/J)}\right\}.
\end{flalign*}The above decomposition clarifies that if the statistic $\bar{Z}$ is impacted by misspecification, such as outliers in the data, then this will ultimately impact our inferences for $\varphi$.

However, the influence of this misspecification can be mitigated by replacing the posterior $\pi(\varphi^2\mid \bar{Z},s^2)$ with the cut posterior that results from updating our knowledge of $\varphi^2$ using the likelihood
for $s^2$ only.  This results
in a posterior for $\varphi^2$ that depends only on
$s^2$, and has the closed form
$$
\pi_\cut(\varphi^2\mid X)\propto \prod_{i=1}^{N}(\varphi_i^2)^{-\frac{J+1}{2}}\exp\left\{-\frac{J\cdot s_i^2}{2\varphi_i^2}\right\}.
$$Joint inferences for $(\zeta,\varphi^2)$ can then be carried out using the cut posterior distribution 
\begin{flalign*}
	\pi_\cut(\zeta,\varphi^2\mid X,Y)	& = \pi_\cut(\varphi\mid X)\pi(\zeta\mid Y,\varphi)=\pi_\cut(\varphi\mid s^2)\pi(\zeta\mid \bar{Z},\varphi).
\end{flalign*}
A change of variable gives the cut posterior for $(\zeta,\varphi)$.  

\begin{figure}[h]
		\centering{\includegraphics[scale=0.3]{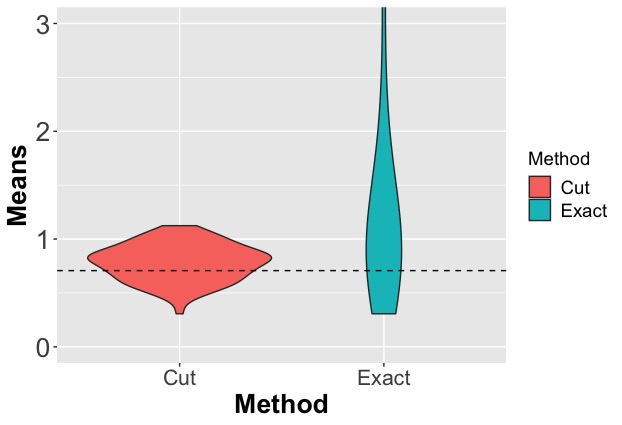}}
		\caption{Violin plots for the cut and exact posterior means of $\varphi_1$.  The true value of $\varphi_1$ is $\sqrt{0.5}$. }\label{fig:cuts}
\end{figure}

To demonstrate the impact of model misspecification on cut and exact Bayesian inference, we compare posterior means for $\varphi_1$ in a repeated sampling experiment. In Figure \ref{fig:cuts}, we present violin plots for the cut and full posterior means of $\varphi_1$ over one-hundred repeated samples with $N=100$ groups, and $J=10$ observations per-group, with $\varphi_i^2=0.5$ for each $i=1,\dots,N$, and $\psi^2=2$. We follow \cite{liu+g22} and set $\beta_1=10$, while simulating the remaining components of $\beta$ from the prior density.  
$\beta_1=10$ is inconsistent with the assumed prior.  The cut posterior delivers more accurate inferences for $\varphi$ than the exact posterior. Even though the misspecification is related to the random effect term $\beta_i$, i.e., the second module, it adversely impacts our inferences for $\varphi$ when using the conventional
posterior, but much less when using cut posterior inferences for $\varphi$. 

\subsection{Obtaining Samples from the cut posterior}

Use of the cut posterior can be attractive in many cases, but 
there are practical challenges of computation to be overcome, as
we now discuss.  
One approach to cut posterior computation is to follow
the sequential decomposition of the cut joint posterior density
in Equation \eqref{cutjoint}, first generating
a $\varphi$ sample from $\pi_{\text{cut}}(\varphi\mid X)$, and then a 
$\zeta$ sample from $\pi(\zeta\mid\varphi,Y)$.  
These draws can rarely be done exactly however.  
The above sequential sampling scheme can in fact be understood
as a modified Gibbs sampling scheme  
where the term $f_2(Y\mid\zeta,\varphi)$ is left out when forming
the full conditional distribution for $\varphi$.  
In more complex situations
than the two-module system of Figure \ref{two-module}, where
there is no clear explicit definition of a cut posterior such as the
one given in Equation 
\eqref{cutjoint}, such a modified Gibbs algorithm can be used to give an implicit
definition of a cut posterior distribution, where terms in the joint
model suspected of misspecification are left out in updating parameters
which might be sensitive to the misspecification.

The connection between cut posterior computation 
and Gibbs sampling might encourage
us to think that we can replace Gibbs steps with Metropolis-within-Gibbs
steps in detailed balance with the modified conditional distributions, in
the case when exact sampling is not possible.  However, as shown
by \cite{plummer15} this is problematic, since  
the stationary distribution for the sampler depends on the proposal
used.  Beyond blocked Gibbs and Metropolis-within-Gibbs approaches, it is 
usually not possible to proceed with any MCMC
algorithm for sampling the joint cut posterior directly.  From Equation \eqref{cutjoint} we see that 
$$\pi_{\text{cut}}(\zeta,\varphi\mid X,Y) \propto 
\frac{f_2(Y\mid\zeta,\varphi)\pi(\zeta\mid\varphi)f_1(X\mid\varphi)\pi(\varphi)}{P(Y\mid\varphi)},$$
and to compute the unnormalized joint cut posterior we need to evaluate the
usually intractable normalizing constant $P(Y\mid \varphi)$.
For the two module case, a 
valid but computationally intensive way to sample the cut posterior distribution
is to first obtain samples $\varphi^{(i)}$, $i=1,\dots, I$, 
from the target $\pi_\cut(\varphi\mid X)$, and then for each value $\varphi^{(i)}$ to run a separate MCMC chain targeting $\pi(\zeta\mid X,\varphi^{(i)})$ to
obtain a single draw $\zeta^{(i)}$.     
In this ``nested MCMC" algorithm each sample of $\zeta$ is drawn
with $\varphi$ held fixed, avoiding the computation of the intractable
$P(Y\mid \varphi)$ term.  \cite{plummer15} discusses both the nested MCMC
method as well as a related tempering approach.

Other proposals for cut model computation include \cite{jacob2020unbiased}, where cut posterior expectations are unbiasedly estimated using coupled 
Markov chains, \cite{liu+g20} who considered a stochastic approximation
Monte Carlo approach, and \cite{pompe+j21} who discuss a posterior
bootstrap method giving valid uncertainty quantification, including for
situations where data sources in different modules are dependent with
each other.  The approach of \cite{pompe+j21} is discussed further
in Section 4.3.  
\cite{yu+ns21} and \cite{carmona+n22} have proposed the use of variational approaches (see \citealp{martin2021approximating} for a brief overview of variational approximation methods) to produce approximations to the cut posterior.

\subsection{When to Cut}
While avoiding impacts of model misspecification is
the motivation for cut methods, it is an open question 
how bad misspecification must be for the use of the cut
posterior to be preferable to the conventional posterior. 
\cite{carmona+n20} framed the issue in terms
of a bias-variance trade-off, saying ``In Cut-model inference, feedback from the suspect module is completely cut. However, [...] if the parameters of a well-specified module are poorly informed by ``local" information then limited information from misspecified modules may allow us to bring the uncertainty down without introducing significant bias.'' Based on the existence of such a hypothetical relationship between bias and variance, several studies have proposed methods to determine when to prefer cut posterior inferences to those based on the exact posterior.  

Conflict checks developed in the literature on Bayesian model
checking (e.g. \citealp{evans+m06,presanis+osd13}) are
one tool which can be used to decide whether cut or full posterior
inferences are preferable.  
The approach suggested by \cite{yu+ns21} for deciding whether or not to cut is based on the relationship between the cut and exact posteriors established in Lemma 1 of \cite{yu+ns21}.  This demonstrates that the Kullback-Leibler (KL) divergence between the exact and cut posterior can be written as a prior-to-posterior divergence in the 
hypothetical situation where the prior information includes the data $X$, 
but $Y$ has yet to be observed.   This result frames the comparison between cut and
full posterior as a prior-data conflict check, for which there are 
well-established methods.   
Denoting the KL-divergence between distributions with densities $p$ and $q$ by $\text{KL}(p,q)$, \cite{yu+ns21} consider the behaviour of the statistic 
$
T=\text{KL}\left\{\pi(\varphi\mid X,Y)||\pi_\cut(\varphi\mid X)\right\},
$ and calibration of such a test statistic can be carried out using the methods in \cite{nott+wee20}, with a Bayesian tail probability produced to determine when there are meaningful differences between the cut and full
posterior distribution. 

Deciding whether or not to cut can also be guided by predictive accuracy, 
as discussed in  \cite{jacob+mhr17} and \cite{carmona+n20,carmona+n22}. 
\cite{jacob+mhr17} consider a decision theoretic framework
for comparing different candidate posterior distributions, 
focusing on generic utility functions defined from logarithmic scoring rules.  
\cite{carmona+n20,carmona+n22} consider using
expected log-pointwise predictive density \citep{vehtari+gg17}, 
approximated using
either cross-validation or WAIC \citep{watanabe13}.
They use their method in the wider context of semi-modular inference, 
discussed next.  

\subsection{Semi-modular Inference}
The binary choice of cutting feedback or not is perhaps artificial, and 
\cite{carmona+n20} proposed the use of a semi-modular posterior (SMP)
that partially cuts feedback.  The SMP allows the analyst to interpolate 
between the cut 
and exact posterior based on a tuning parameter 
chosen by the analyst.  To explain their idea, 
consider once more the two-module system discussed previously.  
The SMP of \cite{carmona+n20} starts by considering a hypothetical
replicate $\widetilde{\zeta}$ of $\zeta$ 
and considers marginal posterior inferences for $\varphi$ based on 
a ``power posterior" (see, e.g. \citealp{grunwald+v17})
$$
\pi_\gamma(\widetilde{\zeta},\varphi \mid X,Y)\propto f_1(X\mid\varphi) f_2(Y\mid\widetilde{\zeta},\varphi)^\gamma \pi(\varphi\mid\zeta)\pi(\zeta).
$$
In this power posterior the influence of the likelihood term for the second 
module has been reduced 
through an ``influence parameter" $\gamma$
with $0\leq \gamma\leq 1$.
The parameters $\widetilde{\zeta}$ can be thought of as introduced auxiliary variables that are used to help learn about $\varphi$. 
By coupling the above distribution
with the conditional distribution for $\zeta$ given $\varphi$ from the conventional
joint posterior we we obtain the modified posterior 
$$
\pi_{\text{smi},\gamma}(\widetilde\zeta,\zeta,\varphi \mid X,Y):=\pi_\gamma(\widetilde\zeta,\varphi\mid X,Y)\pi(\zeta\mid\varphi,Y),
$$
and integrating out $\widetilde{\zeta}$ leads to the SMP  for $(\zeta,\varphi)$, 
which interpolates between the cut ($\gamma=0$) and full ($\gamma=1$) posterior.  The choice of $\gamma$ allows us to regulate the amount of information from the second module to be used 
in inference about $\varphi$. If there is a bias-variance trade off, we may be willing to permit some bias in our inferences for $\varphi$, if this brings a sufficient reduction in uncertainty. 


Sampling from the SMP involves the same difficulties as for the cut posterior, and different values for $\gamma$ need to be considered.
The lack of a straightforward and general sampling scheme for producing draws from $\pi_{\text{smi},\gamma}(\zeta,\varphi \mid X,Y)$ has led to the use of variational approximations to approximate the SMP.  \cite{carmona+n22} generalize the methods in \cite{yu+ns21} to the case of semi-modular inference, consider flexible normalizing flows for the
approximation, and develop approximations to the SMP for all values of the influence parameter $\gamma$ at once, including for multiple influence parameters and cuts.  
However, it remains an open question how best to choose the modularization parameter $\gamma$. Currently, the most common approach to the choice of $\gamma$ is either through the use of conflict checks, as in \cite{chakraborty2023modularized}, or the use of the predictive criteria discussed in \cite{carmona+n20} and \cite{carmona+n22}.  It is not likely that there
will be any uniformly best method, with the preferred approach depending
on the problem at hand.  When the goals are inferential, there is room
for new methods that relate directly to the bias-variance trade-off which
serves as the motivation for semi-modular inference methods.  

While the SMP approach of \cite{carmona+n20} interpolates between the cut and exact posterior distributions using an auxiliary variable, 
several alternative approaches have been suggested in 
\cite{nicholls2022valid}.   The authors also discuss the idea of
valid and order-coherent updating for various SMP proposals.
\cite{chakraborty2023modularized}, and \cite{frazier2023guaranteed}
discussed an alternative SMP based on linear opinion pooling (Stone, 1961). Rather than introduce an auxiliary variable $\widetilde\zeta$ as
in \cite{carmona+n20}, 
\cite{chakraborty2023modularized}, and \cite{frazier2023guaranteed} suggest a linear interpolation between the exact and cut posterior distributions: 
$$
\pi_\gamma(\zeta,\varphi\mid X,Y)=\gamma \pi_\cut(\zeta,\varphi\mid X,Y)+(1-\gamma)\pi(\zeta,\varphi\mid X,Y).
$$However, since the cut posterior takes the form $\pi_\cut(\zeta,\varphi\mid X,Y)=\pi_\cut(\varphi\mid X)\pi(\zeta\mid Y,\varphi)$, the above posterior can be written in terms of a linear interpolation of the marginal $\varphi$ posteriors: 
\begin{equation}\label{eq:linSMP}
	\pi_\gamma(\zeta,\varphi\mid X,Y)=\left\{\gamma \pi_\cut(\varphi\mid X)+(1-\gamma)\pi(\varphi\mid X,Y)\right\}\pi(\zeta\mid Y,\varphi).	
\end{equation}	
Given the construction of this SMP as a linear interpolation, in the remainder we refer to the posterior in equation \eqref{eq:linSMP} as the linear SMP or lin-SMP for short. 

The lin-SMP has three potential benefits over the original SMP of \cite{carmona+n20}. Firstly, the choice of $\gamma$ can be made with respect to the $\varphi$ information only, since the occurrences of $\zeta$ can be marginalised out of the lin-SMP, whereas the choice of $\gamma$ in the SMP of \cite{carmona+n20} is impacted by both $\zeta$ and $\varphi$, due to the nonlinear nature of the power posterior and the introduction of the auxiliary variable $\widetilde\zeta$.  Secondly, lin-SMP does not require us to generate samples from the joint cut posterior in order to choose $\gamma$. Due to its design, samples from $\pi(\varphi\mid X,Y)$ and $\pi_\cut(\varphi\mid X)$ are enough to calculate the value of $\gamma$ (since we can always marginalise out $\zeta$). Lastly, and most importantly, if we choose an optimal value of $\gamma$, the lin-SMP produces inferences that are guaranteed to be more accurate than either the exact or cut posteriors themselves. 
Under a particular asymptotic regime, \cite{frazier2023guaranteed} formally demonstrate the existence of a bias-variance trade off between exact and cut posterior inferences. Using this trade off, the authors demonstrate that an optimal value of $\gamma$ can always be chosen by minimizing a user-defined loss function which encodes the type of statistical robustness that is of interest to the user. When a certain estimate of the optimal $\gamma$ is used, the authors show that the resulting lin-SMP produces inferences that are more accurate than those of the cut posterior or the exact posterior alone.

\section{Parametric projection methods}

The third topic reviewed here is the use of a reference model to generate projected posterior or predictive distributions for a misspecified model. The aim is to approximate predictive inference for the reference model. To implement projection methods, it is only necessary to specify a prior on the reference model, and there is no direct application of Bayes rule in the misspecified model. Many related methods exist, but we do not have enough space to discuss all of them. For example, there are projection methods modelling covariates as random (e.g.~\citealp{lindley68}) and numerous methods for Bayesian model selection under misspecification which assess models based on their predictions, with predictive distributions obtained using conventional Bayesian approaches. For these latter methods prior distributions for parameters in misspecified models must be given, which is conceptually difficult.

We start the discussion of projection methods by considering methods for model choice and variable selection for regression, where an encompassing
model (assumed to be correct) is compared with one or more restricted models
which may be more useful for prediction or scientific understanding.  
This work originates with \cite{goutis+r98} and \cite{dupuis+r03}, with 
a related approach to hypothesis testing not in the regression context 
described in \cite{bernardo99}.
\cite{piironen+pv20} gives a recent review of projection predictive
variable selection.    
There is a largely independent literature on using 
projections with a highly flexible reference model for making
inferences in misspecified parametric models.  
In the case where the data are modelled as independent and identically
distributed conditional on some unknown distribution $P$, a 
nonparametric prior on $P$ can be used (e.g. \citealp{gutierrez-pena+w05}).  
Given some functional $\theta(P)$, it has a
posterior distribution induced by the posterior for $P$, 
and $\theta(P)$ might be defined as a projection onto a misspecified parametric model.  These nonparametric methods will be discussed in Section 4.3, along with applications to cutting feedback methods.  

\subsection{Parametric projections for comparing nested 
generalised linear models}

For model choice in generalised linear models, 
\cite{goutis+r98} considered the following approach.  Let $(y_i,z_i)$, 
$i=1,\dots, n$, be response and covariate pairs to be observed, where
each $y_i$ is a scalar response and $z_i$ is a vector of covariates.  Write
$y$ for the vector of responses, and $z$ for the set of all covariates.  

Write $F$ for the full model including all the covariates, with 
parameter $\theta\in \Theta$, and denote by
$f(y|z,\theta,F)$ the density for $y$ given $z$, $\theta$ and $F$.  
Let $S$ be a model for which the parameter space $\Theta$ is restricted
to $\Theta_S\subset \Theta$;  for example, some components of $\theta$ 
representing regression coefficients might be fixed at
zero to exclude certain covariates.  We denote by $f(y|z,\theta_S,S)$ the 
density of $y$ given
$z$, $\theta_S\in\Theta_S$ and $S$.  
 
Write
\begin{align*}
  d(\theta,\theta_S) & = \int f(y|z,\theta,F)\log \frac{f(y|z,\theta,F)}{f(y|z,\theta_S,S)}\;dy, 
\end{align*}
for the Kullback-Leibler (KL) divergence between 
$f(y|z,\theta,F)$ and $f(y|z,\theta_S,S)$ for given values $\theta\in\Theta$, $\theta_S\in\Theta_S$.    
It will be assumed
that $z$ is fixed at the observed values $z_{\text{obs}}$, although
random $z$ can also be considered.  
Define the KL projection of $\theta$ onto $\Theta_S$ as
\begin{align}
  \theta_S^{\perp}=\theta_S^\perp(\theta)=\arg \min_{\theta_S\in \Theta_S} d(\theta,\theta_S),  \label{projection}
\end{align}
where it is assumed that this projection exists and is unique.  $\theta_S^\perp(\theta)$ is the value of the parameter in the restricted model for which the density of $y$ is closest to that for the full model with parameter $\theta$.  

After specifying a prior density $\pi(\theta)$ for $\theta\in\Theta$, we obtain a 
posterior density
of $\theta$ in model $F$ given the observed responses $y_{\text{obs}}$ and
covariates $z_{\text{obs}}$.  Since $\theta_S^\perp$ is a function
of $\theta$, we can consider the posterior expectation 
\begin{align}
 \delta(F,S) & \coloneqq E(d(\theta_S^\perp,\theta)|y_{\text{obs}},z_{\text{obs}},F),  \label{explanatory}
\end{align}
and \cite{goutis+r98} consider thresholding on this quantity at some $\epsilon>0$
to choose between the hypotheses
$H_0: d(\theta,\theta_S^\perp(\theta))\leq \epsilon$ and $H_1: d(\theta,\theta_S^\perp(\theta))>\epsilon$,  
to decide whether the restricted model $S$ is a 
good enough approximation to $F$.  
A key question is how to choose the threshold $\epsilon$, and 
the authors consider a calibration of the KL divergence
which is useful for this.  A posterior exceedance
probability could be used instead of the expectation 
to make the choice between the hypotheses, but this would involve the choice
of another threshold.  

The generalised linear model assumption is important in the approach
of \cite{goutis+r98}
for addressing the question of existence and uniqueness of the projection, and
for computation.    If the posterior distribution of $\theta$
is approximated by Markov chain Monte Carlo sampling, the posterior
distribution of $\theta_S^\perp$ is approximated by solving
the optimization problem \eqref{projection} draw-by-draw for the 
$\theta$ posterior samples.  This is computationally burdensome, but convenient algorithms exist
for generalised linear models, as well as a closed form expression for
the KL divergence $d(\theta,\theta_S^\perp(\theta))$ once the projection
is obtained.   To explain how to compute the projection, 
consider the situation of a generalised linear model where 
$\theta$ denote regression coefficients in the full model.  For simplicity, 
we assume there is no dispersion parameter;  it turns out that the projection
for the coefficients is the same, no matter what any unknown dispersion
parameter might be.  The projection \eqref{projection} can be 
formulated as 
\begin{align}
  \arg\min_{\theta_S\in\Theta_S} d(\theta,\theta_S) & = 
  \arg\min_{\theta_S\in\Theta_S} \int f(y|\theta,z,F)\log \frac{f(y|\theta,z,F)}{f(y|\theta_S,z,S)} \;dy \nonumber \\
  & = \arg\min_{\theta_S\in\Theta_S} E(-\log f(y|\theta_S,z,S)). \label{mleobjective}
\end{align}
where the expectation in the last line is with respect to $f(y|\theta,z,F)$.  It is easily
seen that minimizing \eqref{mleobjective} is equivalent to obtaining a maximum
likelihood estimate of $\theta_S$ restricted to $\Theta_S$ for 
the case where $y$ is replaced
by its expectation given $\theta$ under $F$.  
We are ``fitting to the fit" of the full model.  Hence in the case
where $\Theta_S$ is defined through a chosen active subset of 
the available covariates, standard software
for obtaining maximum likelihood estimates in generalised
linear models can be used to obtain the projections.   
If $f(y|\theta,z,F)$ does not correspond to a generalised linear model, but
$f(y|\theta_S,z,S)$ does, then the above method for computation
of the projection continues to hold.  

\subsection{Parametric projections in variable selection problems}

Closely related to the method of \cite{goutis+r98} is the variable
selection method developed in \cite{dupuis+r03}.  Now a variable selection
problem is considered in which from a large set of $p$ covariates a subset
must be chosen.  Write $N$ for a null baseline model, such as 
the model with only an intercept.  For any subset model $S$ the relative
loss of explanatory power of $S$ with respect to the full model is 
\begin{align*}
 L(F,S) = & \frac{\delta(F,S)}{\delta(F,N)},
\end{align*}
a number which is between $0$ and $1$. 
\cite{dupuis+r03} suggest choosing $S$ as the simplest model where
the loss of explanatory power is less than some small value chosen by
the user.  If there is more than one such model, then the one with the smallest
loss of explanatory power is selected.   \cite{dupuis+r03} also consider
the case where covariates are random.

\cite{nott+l10} consider a variant of the approach of \cite{dupuis+r03} 
where the
restricted model is defined by an $L^1$ norm constraint on the parameter
of the full model, similar to the lasso \citep{tibshirani96}.  
Once again write $\Theta$ for the parameter space in the full
model and $\Theta_S$ for the parameter space of the restricted 
model $S$.  Suppose
that the parameter $\theta\in\Theta$ is $\theta=(\theta_0,\theta_1,\dots, \theta_p)^\top$, where $\theta_0$ is an intercept and $\theta_j$, $j=1,\dots,p$
are coefficients for the $p$ covariates.   
Then \cite{nott+l10} consider
\begin{align}
 \Theta_S & = \left\{\theta\in\Theta: \sum_{j=1}^p |\theta_j|\leq \lambda \right\}, \label{l1constraint}
\end{align}
where $\lambda>0$ is a tuning parameter.  Computation of the 
projection \eqref{mleobjective} in generalised linear models for this choice of $\Theta_S$ involves fitting a lasso regression where the data $y$
are replaced by their expected values given $\theta$ in the reference model.  
Standard software can be used for this, and sparse projections with many
zero coefficients result. One approach to choosing the tuning parameter 
$\lambda$ uses 
relative loss of explanatory power, similar to \cite{dupuis+r03}.  
\cite{nott+l10} also consider alternative
definitions of $\Theta_S$ inspired by the adaptive lasso \citep{zou06} and
the elastic net \citep{zou+h05}, with corresponding convenient methods
for computing the projections using standard software.    

Instead of the ``draw-by-draw" approach of computing a projected
value of $\theta_S^\perp$ for each of many MCMC posterior draws for
$\theta$, we might consider a projected posterior distribution
putting all its mass on one point (point projection).  
\cite{tran+nl02} 
considered such an approach where, similar to \cite{nott+l10} a 
constraint such as \eqref{l1constraint} can be considered.
\cite{tran+nl02} consider predictive distributions for future observations 
$\widetilde{y}_i$ at covariate values $\widetilde{z}_i$, $i=1,\dots, N$.  
It is possible but not necessary to take $N=n$ and $\widetilde{z}_i=z_i$. 
They minimize a sum of KL divergences between the 
predictive distributions $f(\widetilde{y}_i|\widetilde{z}_i,y_{\text{obs}})$ of $\widetilde{y}_i$ given $\widetilde{z}_i$, $y_{\text{obs}}$ in the
reference model and corresponding predictive distributions with a plug-in form
involving the choice of a point estimate in the projected space.
Computation of the projection involves a criterion similar to  
\eqref{mleobjective}, 
\begin{align}
  \theta_S^\perp & = \arg\min_{\theta_S\in\Theta_S} \sum_{i=1}^N 
  E(-\log f(\widetilde{y}_i|\widetilde{z}_i,\theta_S,S)), \label{pointprojvs}
 \end{align}
 where now the expectation for the $i$th term in the sum is with 
 respect to $f(\widetilde{y}_i|\widetilde{z}_i,y_{\text{obs}})$.  For generalised linear models, projections are computed
 easily as a constrained maximum likelihood problem with data given
 by expected values of $\widetilde{y}_i$ under the reference
 model.  \cite{hahn+c15} derived independently some related methods to
those of \cite{tran+nl02}, with a more explicitly decision-theoretic
perspective.

\cite{piironen+pv20} suggest an interesting projective method in which the 
draw-by-draw and point projection approaches appear as special
cases.  They suggest clustering
of samples from the posterior distribution of the reference model, 
finding a projection for each cluster, and allocating weights to the clusters
proportional to their size.  If the number
of clusters is equal to the number of projected posterior samples, 
we obtain the draw-by-draw projection approach;  if the number of clusters
is $1$, we obtain the point projection method.  
Use of a small
 number of clusters is computationally efficient compared to
 the draw-by-draw method, with significant accuracy improvements relative
 to point projection.  \cite{piironen+pv20} suggest
 $L^1$ and elastic net penalties for obtaining an ordering of the features
 that reduces the number of models to search over.  
 Once the search space has been reduced, they consider projection without
 penalization and assessment of an appropriate complexity to choose 
 for the submodel
 can be done by cross-validation.  They emphasize the importance of doing
 the reference model fitting and feature selections separately for different
 folds of the cross-validation procedure to avoid bias in estimating the
 appropriate subset model size to use.  There have been recent
 efforts to extend the above methods beyond generalised
 linear models.  \cite{piironen+v16} describes variable selection for 
 Gaussian processes, \cite{catalina+bv22} consider model
 choice for generalised linear and additive mixed models,
 \cite{bashir+chj19} consider estimation of sparse precision matrices, 
 \cite{puelz+hc17} consider seemingly unrelated regressions, 
 \cite{kowal22} considers targeted methods
 for prediction and  \cite{peltola18} and \cite{afrabandpey+ppvs20} use projection methods for
 interpretable machine learning.
  
\subsection{Parameter projections with a nonparametric reference model}

We now discuss some of the literature
on projections with a nonparametric reference model. 
\cite{gutierrez-pena+w05} discuss the lack of 
coherence of Bayesian model choice when the prior places probability 
one on a set of densities which does not include the truth.  
They suggest using a flexible nonparametric model for the data, and then 
using parametric models to define predictive distributions for future data
to approximate predictive inference under the nonparametric model.   
Candidate predictive distributions are generally formed by
placing a mixing distribution on the parameter of the parametric model, and
the best predictive distribution maximizes expected utility under the nonparametric reference.  
No Bayesian update for any misspecified parametric
model needs to be considered, and   
different forms of the parametrized predictive density can
be used, corresponding
to various decision problems for
point and interval estimation, predictive inference, model choice and
model averaging.    For point estimation, an optimization problem arises
similar to the one considered earlier for point projection methods in the 
variable selection literature (c.f.~equation \eqref{pointprojvs}).
\cite{walker+g07} consider a related approach to that of 
\cite{gutierrez-pena+w05}, but focusing on predictive densities
formed using a nonparametric
mixing distribution on the parameter of the parametric model and
associated computational issues.

For the choice of the nonparametric reference model, 
convenient choices are
the Bayesian bootstrap \citep{rubin81}, which
is a limiting case of the Dirichlet process \citep{ferguson73}, or Dirichlet
process mixture models   
(e.g. \citealp{lo84,escobar+w95}). 
Several authors have considered the Bayesian bootstrap 
and decision-theoretic arguments to analyse parametric models which are misspecified.  \cite{lyddon+hw19} give an exact interpretation
of the weighted likelihood bootstrap of \cite{newton+r94} using 
the Bayesian bootstrap and a decision-theoretic argument 
which we elaborate on here.  Suppose we have observations 
$y_1,\dots, y_n$ 
assumed to be iid conditional on some unknown 
distribution.   In the Bayesian bootstrap we estimate the distribution 
of the observations nonparametrically, attaching weights 
$\theta=(\theta_1,\dots, \theta_n)^\top$ to the observed sample values.  
The Bayesian bootstrap posterior for $\theta$ is Dirichlet, $\text{Dir}(1_n)$
where $1_n$ denotes an $n$-vector of ones.   
Consider a parametric
model $f(y|\eta)=\prod_{i=1}^n f(y_i|\eta)$.  
For a new observation $\widetilde{y}$, 
the utility for choosing parameter $\eta$ 
when $\widetilde{y}$ is observed is assumed to be $\log f(\widetilde{y}|\eta)$.  
The expected utility under the Bayesian bootstrap conditional on 
a posterior draw for $\theta$ is $\sum_{i=1}^n \theta_i \log f(y_i|\eta)$.
and the corresponding projection of $\theta$ to the parameter space
of the parametric model is  
$$\eta^\perp(\theta)=\arg \max_{\eta} \sum_{i=1}^n \theta_i \log f(y_i|\eta).$$  
 \cite{lyddon+hw19} also
considered generalised Bayes extensions of the above idea.  The procedure
for projecting draws from the posterior of $\theta$ to the parameter 
space of the parametric model is similar to the ``draw-by-draw" projections discussed for the variable selection context.  

An interesting and common application  of projected Bayesian bootstrap methods is to Bayesian propensity score regression for the estimation of average
treatment effects in causal inference.  See for example 
\cite{stephens+nms22} and \cite{saarela+bd16}.  
The cutting feedback methods of the previous section are also used
in this problem
(e.g. \citealp{mccandless+des10,zigler+wyycd13}) and 
\cite{pompe+j21} develop a projective approach to Bayesian
modular inference which is useful here.  
They first explore the frequentist properties of two stage Bayesian
cutting feedback methods, and note
that they do not provide accurate uncertainty quantification in a frequentist
sense.  This problem can also be caused by dependence between
the data sources in two modules, something which occurs in propensity
score regression applications.  
As a remedy, they suggest a
a posterior bootstrap approach \citep{pompe21}.   
Posterior bootstrap samples are projected to parameters
within modules in a two stage approach using utilities defined from
parts of the parametric model for different modules.  A different implementation
of the posterior
bootstrap is required depending on whether datasets
in different modules are dependent or not.  
The posterior
bootstrap discussed in \cite{pompe21} is 
related to the weighted likelihood bootstrap of  
\cite{newton+r94}, the loss-likelihood bootstrap of \cite{lyddon+hw19} and 
another posterior bootstrap proposal of \cite{fong+lh19}.  
It would be simple to use weighted likelihood bootstrap instead of 
posterior bootstrap for projections in the approach of \cite{pompe+j21}, if
specifying a prior in the misspecified parametric model is difficult.
The method 
of \cite{pompe+j21} provides a nice link between the ideas discussed in 
this section and the Bayesian modular inference methods of Section 3.

\section{Discussion and future challenges}

The methods discussed here are undergoing rapid development and 
there are many interesting unanswered questions. 

One of the major obstacles in employing the restricted likelihood approach for robust regression beyond linear models lies in the development of practical computational implementations. While the likelihood-free version of Bayesian Robust Regression (BRL) can theoretically handle complex regression models by relaxing exact conditioning, the conventional statistical approaches to likelihood-free inference don't scale well when many summary statistics are needed.  In the case of M-estimators, each summary statistic corresponds directly to a parameter, and hence marginal adjustment techniques (see \citet{Drovandi2022} and the references therein) where accurate marginal posteriors are estimated with relevant low-dimensional statistics, may be useful.
Methods in machine learning have been developed with the potential to scale to a large number of parameters \citep{cranmer+bl20}.  These methods build a machine learning model such as a neural network from model simulations in order to learn either the likelihood (e.g.\ \citealp{Papamakarios2019}), posterior (e.g.\ \citealp{greenberg+nm19}) or likelihood ratio (e.g.\ \citealp{Thomas2022}).  Some work has been done on extending these methods to handle model misspecification (e.g.\  \citealp{Ward2022,Kelly2023}), although research is on-going.  Applying machine learning likelihood-free approaches for the robust regression problem will be explored in future work.  

An important ongoing issue in modular Bayesian inference is how to define ``modules" and the cut posterior in a general way. A recent pioneering work in this direction is \cite{liu+g22}, who present a possible definition of a ``module" in terms of partitioning of the observables and the graphical structure of the joint model.
Semi-modular inference is another active current research topic for which
there are important unresolved questions.
\cite{frazier2023guaranteed} give statistical guarantees for the lin-SMP approach, but there is no guarantee that lin-SMP delivers the most
useful approximation across all classes of posterior interpolations.  It is worthy of investigation whether the accuracy guarantees of the lin-SMP method also apply to the original SMP of \cite{carmona+n20}, and the variational extensions in \cite{carmona+n22}.  

Finally, for the projection techniques discussed in Section 4, perhaps the main
challenge is computation.  Most work on these methods has been
confined to situations where the projection involves an exponential family
model such as a generalised linear model, where the optimization problems
to be solved are well understood and can be performed with standard software.  
Dealing with misspecification, however, is a problem faced often in complex
situations, which means it is important to find more general approaches
to implementation. 

\vspace{5mm}

\section*{Disclosure statement}
The authors are not aware of any affiliations, memberships, funding, or financial holdings that
might be perceived as affecting the objectivity of this review. 

\section*{Acknowledgements}

D.J.N. is affiliated with the Institute of Operations Research and Analytics, National University of Singapore.
The work of C.D.\ was supported by an Australian Research Council Future Fellowship (FT210100260). The work of D.T.F. was supported by an Australian Research Council Early Career Fellowship (DE200101070).

\bibliographystyle{chicago}
\bibliography{Bayesian-misspecified}

\end{document}